\documentclass[useAMS,usenatbib,usegraphicx]{mn2e}
\usepackage{times}

\title[{\it Kepler} results on compact pulsators II: KIC\,010139564]
{First {\it Kepler} results on compact pulsators II:  KIC\,010139564, a new pulsating subdwarf B (V361~Hya) star with an additional low-frequency mode}

\author[S. D. Kawaler et al.]
{S. D. Kawaler$^1$,
M. D. Reed$^2$,
A. C. Quint$^2$,
R. H. \O stensen$^3$,
R. Silvotti$^4$
\newauthor
A. S. Baran$^{1,5}$,
S. Charpinet$^6$,
S. Bloemen$^3$,
D. W. Kurtz$^7$,
J. Telting$^8$,
G. Handler$^9$,
\newauthor
H. Kjeldsen$^{10}$,
J. Christensen-Dalsgaard$^{10}$,
W. J. Borucki$^{11}$,
D. G. Koch$^{11}$
\newauthor
and
\newauthor
J. Robinson$^{12}$\\
$^{1}$Department of Physics and Astronomy, Iowa State University, Ames, IA 50011 USA\\
$^{2}$Department of Physics, Astronomy and Materials Science,
 Missouri State University, 901 S. National, Springfield, MO 65897 USA\\$^{3}$Instituut voor Sterrenkunde, K.U.~Leuven, Celestijnenlaan 200D, 3001 Leuven, Belgium\\
$^{4}$INAF-Osservatorio Astronomico di Torino, Strada dell'Osservatorio 20, 10025 Pino Torinese, Italy\\
$^{5}$Krakow Pedagogical University, ul. Podchor\c{a}\.{z}ych 2,30-084 Krak\'{o}w, Poland\\
$^{6}$Laboratoire d'Astrophysique de Toulouse-Tarbes, Universit\'e de Toulouse, CNRS, 14 Av. E. Belin, 31400 Toulouse, France\\
$^{7}$Jeremiah Horrocks Institute of Astrophysics, University of Central Lancashire, Preston, PR1 2HE, UK\\
$^{8}$Nordic Optical Telescope, 38700 Santa Cruze de La Palma, Spain\\
$^{9}$Institute f\"ur Astronomie, Universit\"at Wien, T\"urkenschanzstrasse 17, 1180 Wien, Austria\\
$^{10}$Department of Physics and Astronomy, Aarhus University, DK-8000 Aarhus C, Denmark\\
$^{11}$NASA Ames Research Center, MS 244-30, Moffett Field, CA  94035, USA\\
$^{12}$Alpha Control, USSC, Houston, TX, USA
}

\begin{document}

\date{post-review 2: 12 August 2010}
\pagerange{\pageref{firstpage}--\pageref{lastpage}} \pubyear{2010}

\maketitle

\label{firstpage}

\begin{abstract}
We present the discovery of nonradial pulsations in a hot subdwarf B star based on 30.5
days of nearly continuous time-series photometry using the \emph{Kepler} spacecraft.  KIC\,010139564 
is found to be a short--period pulsator of the V361\,Hya (EC\,14026) class with more than 10 
independent
pulsation modes whose periods range from 130 to 190 seconds. It also shows  one periodicity at  a period of 
3165 seconds.  If this periodicity is a high order $g$-mode, then this star may be the hottest member of the
hybrid DW\,Lyn stars.  In addition to the resolved pulsation 
frequencies, additional periodic variations in the light curve suggest that a significant number of 
additional
pulsation frequencies may be present. 
The long duration of the run, the extremely high duty cycle, and 
the well--behaved noise properties allow us to explore the stability of the periodic variations, and to 
place strong 
constraints on how many of them are independent stellar oscillation modes.   
We find that most of
the identified periodicities are indeed stable in phase and amplitude, suggesting a rotation period
of 2-3 weeks for this star, but further observations are needed to confirm this suspicion.
\end{abstract}

\begin{keywords}
Stars: oscillations -- stars: variables --
Stars: subdwarfs
\end{keywords}

\section{Introduction}

Subdwarf B (sdB) stars are highly evolved low mass stars that have survived core helium
ignition (presumably through the core helium flash) and now populate the extreme blue
end of the horizontal branch. \citet{Heber09} reviews the general properties of these stars.
 Generally, they are expected to have a mass of 
$\approx$0.5 M$_\odot$. Their hot surfaces \citep[$T_{\rm eff} \approx$ 22,000K to 40,000K -][]{heber,saf94} 
indicate
that they retain a thin ($< 10^{-3}$M$_\odot$) hydrogen surface layer \citep{dorm93}.  While the general trends of mass and $T_{\rm eff}$ on the horizontal branch are well understood, the origin of the hot sdB stars remains unsettled.  One plausible mechanism by which stars lose all but a small fraction of their hydrogen envelope has been attributed to binary evolution and mass transfer 
\citep{han02,han03}.

The discovery of multiperiodic pulsating sdB stars \citep{kill97} opened up these stars to asteroseismic probing, promising to give us a clearer picture of their current internal structure and, thereby, their origins.   These pulsators, now known as V361~Hya stars (and as EC\,14026 stars, but often referred to as short period sdBV stars), show multiperiodic pulsations with periods ranging from 1.5 to 10 minutes, with most in the shorter period part of that range.  They typically have pulsation amplitudes ($\Delta I / I$) of 1 percent
and less, with detailed studies revealing a few to dozens of frequencies \citep{reed07a}. 
These periodicities 
represent nonradial $p$-modes of low order.

Soon after their discovery, theoretical work by \citet{charp96, char01} and collaborators 
identified the pulsation driving mechanism as cyclical ionization of iron.  Furthermore, they showed that for driving to occur, radiative levitation needed to enhance the abundance of iron within the driving region.  Their models demonstrate that equilibrium computation of diffusion effects can indeed produce the iron enhancement needed for driving stars at the effective temperatures (and gravities) of the V361~Hya stars; see \citet{char01} and \citet{font06} for an overview of driving in sdB stars.

There is a second, related class of pulsating sdB stars, discovered by \citet{grn03}.  These stars, known as the V1093\,Her stars (or PG\,1716 stars, or long period sdBV stars), show longer
period pulsations ranging from 45 to 120 minutes or more, and
typically have semi--amplitudes lower than 0.2 per~cent \citep{royjenam}.  
They are
cooler than the V361\,Hya stars, though there is some overlap, and
they are most likely $g$-mode pulsators \citep{font03}.
A third type
are hybrid (DW~Lyn) pulsators which show both long and short period variations 
\citep{baran05, schuh06}.  In most cases, the DW\,Lyn stars show much larger pulsation amplitude
in the higher-frequency modes, by factors ranging from 5 to 20 or more.


As sdBV stars
often pulsate with tens of frequencies, with each frequency probing a
slightly different region of the interior, successful application of
asteroseismology has great potential for probing the interiors of these stars.
Asteroseismology of pulsating sdB stars can provide determinations of their global properties such as mass and radius.  Matching the pulsation frequencies  with model values can further constrain their internal structure in terms of the thickness of their surface hydrogen layer and, perhaps, the size of the helium-burning core  \citep{charp07} and internal rotation profiles \citep{kawhos,vang08,charp08}.  The mere presence of pulsations allows testing models of radiative levitation and diffusion in these stars as well \citep{charp09}.  Taken together with spectroscopic analysis, the pulsating sdBV stars may indeed provide the long-sought window into these stars that can reveal their origins as either single stars, or the product of binary evolution \citep{han02, han03,  hu08, hu09}.

Continuing challenges for asteroseismology lie in the stability of the observed periodicities, and in the identification of the degree and order of the observed frequencies.  They are generally complicated pulsators, with closely-spaced periodicities that result in beating on time scales of hours to days or even longer.  Furthermore, the intrinsic amplitude of certain pulsation frequencies appears to change on the time scales of days to weeks \citep{kilk10}.  To make precise models of these stars, as well as to clearly determine rotation periods of days to months through multiplet splitting, we need to know which periodicities are intrinsic, and which show up in the Fourier transform because of amplitude (or phase) modulation. 

These effects are difficult to sort out with ground-based data because of diurnal aliasing, transparency variations and other atmospheric effects.  As a result, disentangling the presence of closely-spaced (in frequency) periodicities from intrinsic amplitude variation has been a challenge for many of the known sdBV stars. We are still uncertain whether intrinsic amplitude modulation is a ubiquitous feature of sdBV stars.  Long, high duty cycle observations of sdBV stars should help us answer these questions.

For their potential as probes of an important and poorly understood phase of stellar evolution, sdBV stars were included as targets for the asteroseismology survey phase of the {\em Kepler Mission}.  
The \emph{Kepler} science
goals, mission design, and overall performance are reviewed
by \citet{bor10} and \citet{koch10}. Asteroseismology for \emph{Kepler}
is being conducted through the Kepler Asteroseismic Science Consortium
(KASC)\footnote{The Kepler Asteroseismic Investigation at the time of this work was managed by 
R. Gilliland, T. Brown, J. Christensen-Dalsgaard, and H. Kjeldsen.}. 
These
data were obtained as part of the survey mode, where short-cadence
targets are observed for 30 days each during  the first year of the mission

We observed a number of candidate sdB stars during this phase, and have discovered several pulsators.  The target selection process for sdB stars, and results of the first 6 months of the survey phase, are described in  \citet{roy10P1} (hereafter Paper I).  In this paper, we present the discovery of a new pulsating sdB star, Kepler Input Catalog (KIC)~010139564, at $\rmn{RA}(2000)=19^{\rmn{h}} 24^{\rmn{m}} 58\fs15$, 
$\rmn{Dec.}~(2000)=47\degr 7\arcmin 53\farcs 6$, that shows short-period pulsations and also shows evidence of being a hybrid (DW~Lyn) star. Our results suggest that many of the periodicities are quite stable in frequency, amplitude, and phase.  Some of the periodicities are not as yet fully resolved in frequency, but we anticipate that further data will be able to determine whether their complexities are caused by intrinsic pulsations that are closely spaced in frequency, or are caused by amplitude and/or phase variations. 

The V316\,Hya stars, such as KIC\,010139564, are in general much rarer than the longer-period V1093\,Her stars.  Through the first 3/4 of the {\it Kepler} asteroseismology survey phase, KIC\,010139564 is the only V361\,Hya star found in the {\it Kepler} field. 
For first results from the  \emph{Kepler}  survey phase for the long period sdBV stars see \citet{lpsdBVP1} (Paper III), with asteroseismic determinations of the properties of one of those stars presented in \citet{kpd1943} (Paper IV).  First results for long period sdB pulsators in binary systems are presented by \citet{sdbbinary} (Paper V).  A bright, eclipsing sdB+dM system with a primary that is pulsating with both $p$- and $g$- modes was also discovered in the {\em Kepler} commissioning data
\citep{ostensen2010c}.

\section{Observations}
KIC\,010139564 has a {\it Kepler} magnitude ({\em Kp}) of 16.13 magnitude.  
The {\em Kp} magnitude is a broadband magnitude reflecting the colour response of the {\em Kepler} photometer, with a central wavelength roughly equivalent to the $R$-band and a passband effectively spanning 437 to 897\,nm \citep{koch10}.
This is a spectroscopically confirmed sdB star (Paper I) 
with $T_{\rm eff}$\,=\,$32,500\pm200$K and $\log g$\,=\,$5.81\pm0.04$, and with a very small surface helium abundance:  $\log$(He/H)\,=\,-2.8.
The data were obtained  between BJD~2455002
and 2455033 (20 June - 20 July, 2009).
During these $\approx 30.5$ days of observations, a total of 45,210 science
images were scheduled. 
However, spacecraft-related events occasionally result in loss and subsequent reacquisition of fine guidance, which results in individual or short sequences of images being unusable.  In addition, cosmic ray interference and software glitches can result in longer down-times (as the spacecraft enters ``safe mode'') for which no data are available.  During this run, 3,686 images were either not obtained or were unusable.
The photometric data obtained for this target
are short cadence data with a time between points of 58.85~s and a duty cycle (within the 58.85~s window) of 92 percent \citep{GilSC}.
Because of the gaps during the run, and other small deviations in the sampling rate, we determined an effective Nyquist frequency of  8496.20~$\mu$Hz by the symmetry in the power spectrum at frequencies surrounding this value.

Raw and corrected photon counts are provided by the data pipeline. The ``corrected'' flux values included a preliminary estimate of the contamination from a nearby (assumed constant) star that contributes light to the 4 arc second square pixel in the focal plane.  This correction subtracts a constant flux, which increases the resulting fractional amplitude of variations for the target star.
For this data set, the estimated contamination fraction is 0.6 (that is 40 per~cent of the flux from the target and 60 per~cent from background or foreground stars).  
At this time we do not have data on the contaminating source, but we recognize that it could perhaps be variable, and contribute to peaks in the power spectrum.
We expect that the contamination estimate (and the properties of the source) will be determined in the future, but  for this work we report results for the raw fluxes. 

\begin{figure}
\includegraphics[width=84mm]{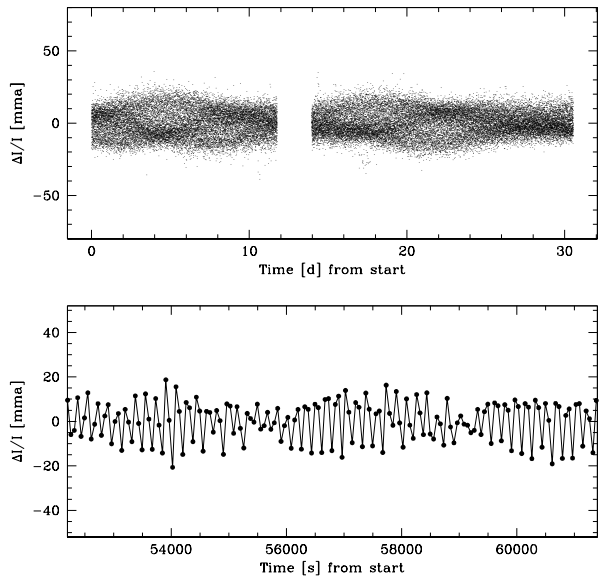}
\caption{Light curve of KIC\,010139564.  The horizontal axis is time (in days in the upper panel, seconds in the lower panel), and the vertical axis shows the differential flux variation in mma (parts per thousand).  The upper panel shows the complete light curve, while the lower panel shows a shorter segment of the light curve from the first day of observation, illustrating the pulsations that are easily seen in the time series.} \label{fig01}
\end{figure}

\emph{Kepler} data on this star were provided with times in barycentric corrected
Julian days, and flux.
Variations of the normalized  ``raw'' flux from the mean (of 1) are shown in the light curve in Fig.~\ref{fig01}.  
Note the gap during the data that corresponds to a 
safing event.  Amplitudes are
given as parts-per-thousand (or milli-modulation amplitude, mma), with 10~mma
corresponding to 1.0 per~cent. Since some contamination by background sources is included (see earlier discussion) the intrinsic pulsation amplitude is greater than this.  Future processing (with an accurate estimate of contamination) should provide a more realistic absolute amplitude.
The short period variations are clearly seen in the light curve - in fact, much of the observed width of the light curve in the top panel is a result of the pulsational modulation.  This is apparent when examining the lower panel of Fig.~\ref{fig01}, which shows a periodic variation of approximately 175 seconds, with a peak-to-peak variation of 30~mma.  

Note that at the time of this writing the pipeline for
reducing \emph{Kepler} short cadence data is still being fine-tuned
and tested. As a result, it is possible that some frequencies (and likely that the pulsation amplitudes)
could be affected by the current \emph{Kepler} reduction
pipeline with differing results in subsequent data releases as the
process is improved.

\section{Time-Series Analysis}

The beauty of data from the \emph{Kepler} spacecraft is that we get nearly continuous
data acquired from a single instrument in a uniform manner. As 
such, analysis can proceed in a straightforward way.  We begin with a standard Fourier transform / 
temporal spectrum of the entire light curve up to the Nyquist frequency of $8496.2\, \mu$Hz.  The main frequency range showing oscillations is from 5000 to 7000~$\mu$Hz.  This region is shown in the top panel of Fig. \ref{fig02}.

One point of concern for short period sdBV stars is that the sampling rate places the Nyquist frequency at $8496\,\mu$Hz, meaning that periods shorter than 117.7 seconds lie beyond the Nyquist frequency.  This  can be an issue as a many sdBV stars have pulsation frequencies beyond that
range \citep{roysdbcat}. Shorter periods will be mirrored across
the Nyquist frequency into the lower frequency domain, creating spurious
frequencies of intrinsic variability to the star. In addition, since the oscillation frequencies are relatively close to the Nyquist frequency, and data are obtained by integration over the short-cadence interval of 58.8 seconds, the amplitudes that appear in the temporal spectrum are reduced below the intrinsic amplitude.  This reduction factor is approximately 25 per~cent for the shortest period frequency, and 13 per~cent for the periodicities near 170-180 seconds. 

Also, though the data we examined were {\em Kepler} short cadence (SC) data sampled every 58.84 seconds, the spacecraft also supplies long-cadence (LC) data with intervals of 30 times the SC rate.  Some part of the process by which the LC data are output affects the SC data by creating artefact peaks at multiples of the LC frequency of $n\times 566.427\,\mu$Hz \citep{GilSC}. This artefact is not well characterized, but it is often most noticeable at higher harmonics that can impact the analysis of the short period sdBV stars.   We examined data on many stars obtained during this same observing period as KIC\,010139564, and detected common artefact peaks at 3965, 4531, 5098, and 5664~$\mu$Hz.  Therefore, we do not further consider periodicities at these frequencies in KIC\,010139564.

\subsection{Fitting periodicities revealed by the temporal spectrum}

We estimate the significance of peaks in the temporal spectrum as in \citet{breger93}:  in frequency ranges near the range of interest, but that do not contain obvious significant peaks, we find the mean value of the amplitude spectrum (the square root of the power).  In this case, in the frequency range of interest, the mean value of the temporal spectrum  is approximately 37 parts per million, or an amplitude of 0.037~mma.  Defining that mean as  $\sigma$, we adopt an amplitude threshold for significance of 4-$\sigma$. We note that this is just one option in determining a significance threshold; a more conservative approach is to define $\sigma$ as the mean amplitude of {\em peaks} in the temporal spectrum, since it is the peaks that are used in identifying candidate periodicities.  However, here we will use the mean of the temporal spectrum, rather than the peaks, to remain consistent with prior work.  We therefore adopt a 4~$\sigma$ detection limit  of 0.15~mma in the range of interest in KIC\,010139564.  

Beginning with the highest peaks in the temporal spectrum we successively fit each periodicity (via non-linear least squares) in the time domain, and then removed each in the time domain, progressing down to the 4-$\sigma$ detection limit.  In this procedure, we did not fit peaks that were within the frequency resolution limit of previously removed periodicities, nor did we fit peaks that were not visible in the original power spectrum.  The frequency resolution of the data is formally $0.38 \mu$Hz, which corresponds to the reciprocal of the time span of the data.

\begin{figure}
\includegraphics[width=84mm]{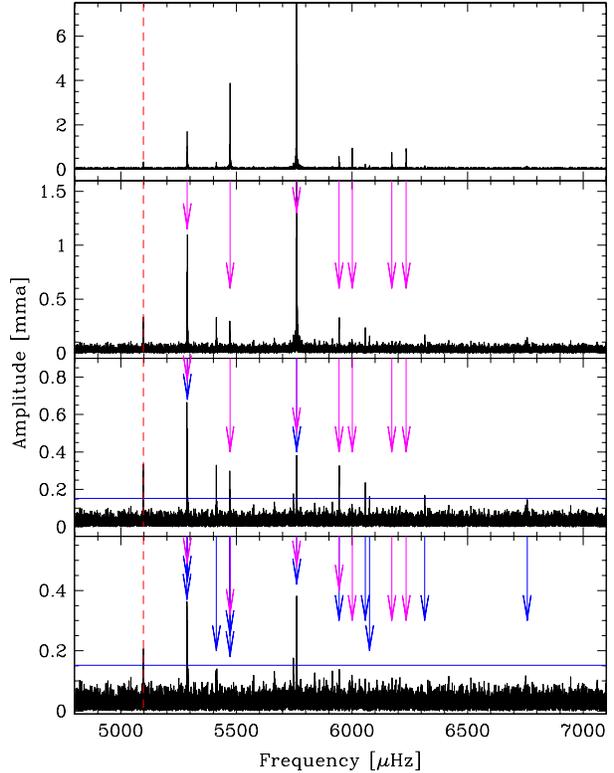}

\caption{Temporal spectrum of KIC\,010139564. Top panel is the temporal spectrum
of KIC\,010139564 (the highest peak, at 10~mma, is truncated); the second panel is for the data with seven periodicities (indicated
by magenta arrows) removed via prewhitening.
Two additional periodicities (indicated in blue) were removed for the third panel. The bottom
panel is the temporal spectrum of the residuals after all frequencies in Table 1 have been removed.
The artefact at 5097~$\mu$Hz is indicated by a dashed (red) vertical line.  The horizontal blue line shows the 4-$\sigma$ detection limit.
Note that the vertical scale changes with each panel.} \label{fig02}
\end{figure}

The prewhitening procedure is illustrated in Fig.~\ref{fig02}. In this figure, the
top panel shows the original temporal spectrum covering the region where most of the periodicities were found. 
The middle panels
show the residuals after removal of seven and then nine high-amplitude frequencies, and the bottom
panel has 22 frequencies removed (including two frequencies outside of the illustrated range, and one at the position of the instrumental artefact at 5098~$\mu$Hz. The dashed (red) line indicates
an artefact frequency mentioned above, the solid horizontal (blue)
line indicates the 4-$\sigma$ detection limit, and the arrows
indicate identified and fit periodicities. The arrows are color-coded (in the 
electronic version, shades of gray in the printed version) to indicate
the stages of the fitting process. The magenta (lightest) arrows indicate the seven frequencies
which were most readily resolved and fit, blue indicates other frequencies which were apparent after the prewhitening
of the initial seven periodicities.  Some of these are extremely close to the initial seven, but others are quite well separated.

In this star, there are at least four
periodicities with amplitudes which would make them easily
detectable from the ground. However, the bulk of the pulsations have
amplitudes below usual single-site Earth-based detection limits for a 16th magnitude star 
observed with a 1-2m class telescope.
The solution to our fit, listing frequencies, periods, and amplitudes is provided
in Table~\ref{tab01}.
\begin{table*}
\begin{minipage}{126mm}
\caption{Periodicities in KIC\,010139564
Formal least-squares errors in parentheses.\label{tab01}}
\begin{tabular}{lccccl}
\hline
ID & Frequency [$\mu$Hz] & Period [s] & Amplitude [mma] & Stability test & Comment \\
\hline




f0  & 315.960 (30) & 3164.96 (30) & 0.435 (67) \\
f1  &5286.046 (30)   &  189.1773 (07) & 0.698 (66) \\
f2  &5287.298 (08)   &  189.1325 (03) & 1.847 (68) & *  \\
f3  &5287.798 (12)   &  189.1146 (04) & 1.173 (65) & & alias of f8+f10 ?\\
f4  &5413.385 (40)   &  184.7273 (13) & 0.329 (68) \\
f5  &5471.184 (76)   &  182.7758 (25) & 0.187 (65)  & & alias of f8+f9 \\
f6  &5471.930 (46)   &  182.7509 (15) & 0.297 (64)  & & alias of 2$\times$f8 \\
f7  &5472.864 (03)   &  182.7197 (01) & 3.891 (66) & *  \\
f8  &5760.229 (01)   &  173.60420 (04) & 9.998 (64)& * \\
f9  &5760.980 (04)   &  173.5816 (01) & 3.314 (64) \\
f10  &5944.369 (22)   &  168.2264 (06) & 0.623 (66)  & * \\
f11  &5944.982 (39)   &  168.2091 (11) & 0.349 (66) & & alias of f2+f8 ?\\
f12  &6001.343 (14)   &  166.6294 (04) & 0.956 (66) & * \\
f13  &6057.601 (55)   &  165.0818 (15) & 0.236 (66) \\
f14  &6076.316 (80)   &  164.5734 (22) & 0.162 (64) \\
f15  &6172.424 (17)   &  162.0109 (04) & 0.762 (66) & * \\
f16  &6234.673 (14)   &  160.3933 (04) & 0.941 (67) & * \\
f17 &6315.043 (77)   &  158.3520 (19) & 0.169 (64) \\
f18 &6758.208 (90)   &  147.9682 (20) & 0.145 (64) \\
f19 &7633.647 (75)   &  130.9990 (13) & 0.175 (65) \\
\hline
\end{tabular}
\end{minipage}
\end{table*}

\begin{figure*}
\begin{minipage}{126mm}
\includegraphics[width=126mm]{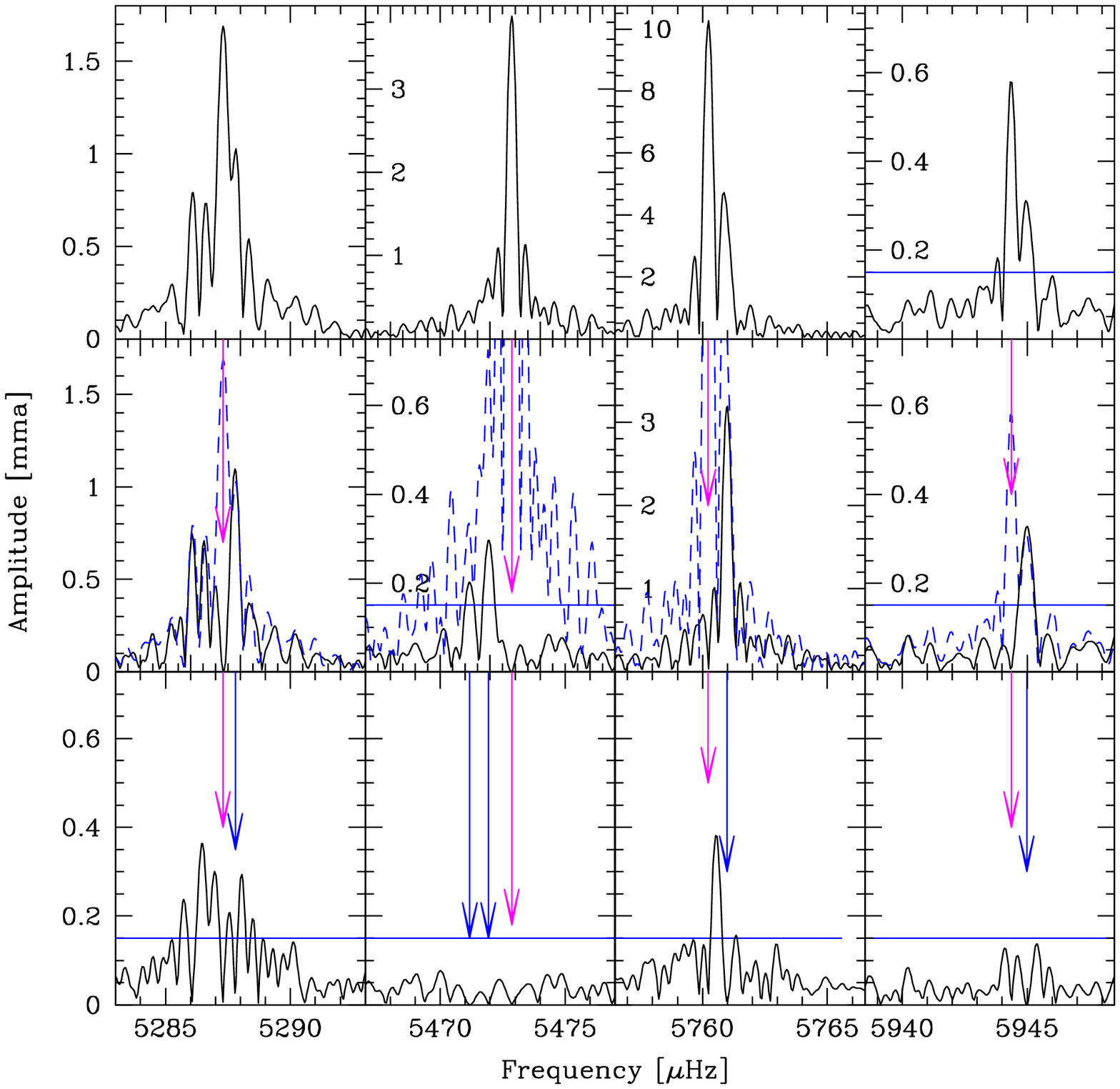}
\caption{Enlarged sections of the temporal spectrum of KIC\,010139564.
The top panels are original temporal spectrum, and the middle
and bottom panels are prewhitened by the indicated frequencies. 
The 4-$\sigma$ detection
limit is indicated by the horizontal (blue) line and the dashed
(blue) line in the middle panels is the original FT, to show
how the high-amplitude peaks affect the others in that region.
Note that the vertical scale changes with each panel.} \label{fig03}
\end{minipage}
\end{figure*}

While the pulsations span over $1400\,\mu$Hz, most frequencies are
within a range of approximately $1000\,\mu$Hz. There are crowded regions where
the periodicities are closely spaced or even unresolved at the current resolution of 0.38~$\mu$Hz.
The crowding is illustrated in Fig.~\ref{fig03}. As in Fig.~\ref{fig02} the horizontal
(blue) line is the 4-$\sigma$ detection limit and the arrows indicate the positions of various 
periodicities that we included in the fit. Each vertical panel shows a $10\,\mu$Hz region with the top panels
showing the original FT, the middle panels are prewhitened by the
highest amplitude frequency in the region. The dashed (magenta) line
indicates the original FT in this panel and illustrates the difficulties
with removing additional frequencies. The bottom panels show the regions
after prewhitening at the indicated frequencies. The bottom panels are plotted on the same vertical
scale, but the others have scales to optimize the visibility of
peaks. Despite having 30.5 days of nearly continuous data, there remain
at least three areas of unresolved periodicities.

\subsection{Combination frequencies, harmonics, and aliasing across the Nyquist frequency}

Many pulsating sdB stars show frequencies that are higher than the Nyquist frequency present in the {\em Kepler} SC data.  Furthermore, large-amplitude pulsators often show harmonics of the main pulsation peaks which can have high frequencies
\footnote{For example,  the sdB pulsator PG~1325+101 \citep{Silv06, charp06} has $T_{\rm eff}$ and $\log g$ that is very similar to KIC\,010139564 and is useful comparison.  PG~1325+101 has a dominant pulsation mode at 7255.5~$\mu$Hz and its harmonic at 14,511~$\mu$Hz.  It also shows other periodicities ranging from 5960 to 10,545$\mu$Hz.}.
In this star, the first harmonic of f8 should be at a frequency of 11520.26~$\mu$Hz, which is 3024.26~$\mu$Hz above the Nyquist frequency.  Thus an alias of the harmonic would appear at $ f_{\rm Ny} - 3024.26$~$\mu$Hz, or 5471.94~$\mu$Hz.  There is indeed a peak in Table\ref{tab01} at that frequency.  Thus we identify f6 as the (aliased) first harmonic of f8.

Similarly, high-amplitude sdB pulsators sometimes show periodicities at linear combinations of strong pulsation modes.  For the periodicities in this star, frequency sums will also be above the Nyquist frequency.  We see one example of this (noted in Table
\ref{tab01}): the sum of f8 and f9 lies above the Nyquist, and its alias peak lies precisely (well within our errors) where f5 is found.  Thus we conclude that f5 and f6 result from non-linear behavior of f8 and f9.  Similar ties to the dominant f8 mode link f3 and f10 (i.e. the alias of ${\rm f}3 + {\rm f}8$ lies at precisely f10) and f2 and f11 (the alias of ${\rm f}2 + {\rm f}8$ lies within 0.11~$\mu$Hz of f11).  Because of the rather low Nyquist frequency for this star, and the complex structure of the temporal spectrum in the region of f2 and f3, it is difficult to determine, with confidence, which periodicities represent oscillation modes, which are combination frequencies, and which might be aliases of ``true'' peaks at higher frequencies.  We hope that continued {\em Kepler} observations will resolve some of these ambiguities, but higher--cadence ground--based photometry will also be needed.

To summarize, we find 20 individual periodicities through a temporal spectrum / least squares fitting procedure, with 19 distributed nonuniformly across the range of frequencies expected for an short-period sdBV star.  Of those, at least two and perhaps four or more represent harmonics and linear combinations of the stronger periodicities, aliased across the Nyquist frequency.  We find a single low-frequency periodicity at $316 \mu$Hz that is within the frequency range seen in the long-period sdBV stars.  Thus, KIC~010139654 appears to be a hybrid sdB pulsator 
(though there is a small chance that this periodicity could be caused by the contaminating source). 
In the next section, we examine the periodicities further to see if we can determine if they are all independent oscillation modes,  or if some of the closely-spaced peaks are the signature of slow amplitude and/or phase modulation.

\section{Discussion}
\subsection{Constraints on the pulsation modes}

For successful asteroseismic analysis, the observations need to provide determination of (or useful constraints  on) the modes of oscillation represented by the observed periodicities.  This includes a determination of the degree $l$ of the oscillation mode.  The mode frequencies themselves are described by 
spherical harmonics determined by the quantum numbers $l$, the mode order $n$, and (if rotation is present) the azimuthal quantum number $m$.  When rotation is present, it can readily reveal the value of $l$ if all $2l+1$ multiplets are present.  From a stellar seismology perspective, this was demonstrated quite clearly in early ground-based networked observations of pulsating white dwarfs \citep[i.e.][]{wing91} but has been used extensively for other compact pulsators as well \citep[i.e.][]{me2,simon}.  Time--series photometry rarely uncovers modes with $\ell > 2$, so the multiplet structure should be well defined - unless the rotational splitting is larger than the spacing between modes with successive $n$, as can be the case in $\delta$~Scuti stars \citep[i.e.][]{temp97}.  

In sdB stars, we expect that their rotation rates are slow enough \citep{heber99,heber00} that such overlaps should not be a problem \citep{vang08} except in some extreme cases \citep{kawhos}. For slow rotators, rotationally-split multiplets should be nearly equally spaced in
frequency. Yet such multiplets are seldom seen in sdBV stars.  This is where \emph{Kepler} data should be most useful, as it
should clearly resolve rotationally split multiplets in sdB stars even if they have rotation periods of months. However,
data from the survey phase spanned only 30.5 days and so to resolve any multiplets, the rotation period would need to be shorter than that. 

We find no obvious multiplets in the periodicities of KIC\,010139564.  With 
a formal resolution of 0.38~$\mu$Hz, the smallest 
barely-resolved spacings that we find are 0.5 to 0.6~$\mu$Hz.  Hence we can state that the 
data are consistent with a rotation period corresponding 
to $P_{\rm rot}$ about 20 days or longer ($1/P \approx 0.6\,\mu$Hz). 
That said, the amplitude of
 $m \ne 0$ multiplets does depend on the inclination of the rotation axis to the line of sight, 
so if the axis is pointing towards the Earth the splitting would be at very small amplitude 
compared to the central peak. There are regions where the pulsations
are clearly unresolved and these regions are the most likely places to 
discern multiplets. We will need to wait for additional
\emph{Kepler} data to see if these regions resolve into multiplets.


Without mode identifications via multiplets, 
we can, as a weak overall constraint, 
appeal to another property of sdB oscillations -- the density of periodicities within a frequency range. Matching the observed $T_{\rm eff}$ and $\log g$ with a stellar model allows us to estimate the number of available mode frequencies within the observed range. 
In resolved sdBV stars, we sometimes observe many more pulsation frequencies than $l = 0$, 1, and 2 can provide, independent of the number of inferred $m \neq 0$ modes.
If higher $l$ modes contribute to the observed number of periodicities, they must have a
higher intrinsic amplitude than the lower $l$ modes  because of the large degree of
geometric cancellation \citep{dziem77,charp05,me1}.

Models appropriate for KIC~010139654, such as those in \citet{charp06} for PG~1325+101 (which has similar values for $T_{\rm eff}$ and $\log g$), show roughly one overtone ($n$) per $l$ degree per 850~$\mu$Hz.
Under normal observing conditions, this would supply 
three to four modes
per 1000~$\mu$Hz ($l=0, 1, 2$) 
without invoking higher $\ell$ values and if no multiplet structure is present.  

For KIC~010139564, the peaks f1 to f17 in Table~\ref{tab01}
span slightly more than $1000\,\mu$Hz.
If the closely-spaced periodicities are considered as 
multiplets of one overtone, then 10 
modes remain, which cannot be accommodated by 
low $l$ modes \footnote{If we allow for $l=3$ and $l=4$ modes, recognizing that $l=4$ modes suffer less geometric cancelation than $\ell =3$ modes for most $m$ values \citep{charp05,me1}, the number of modes per 1000~$\mu$Hz rises to 6. } 


The amplitudes of the \emph{Kepler} periodicities in Table~\ref{tab01} span nearly a factor of 70, so that it is possible that the lower-amplitude periodicities might have $\ell > 2$.  If further {\em Kepler} observations span one year, and the modes remain stable, the noise level will be reduced by a factor of 3 or more, and if higher $l$ modes are present we might hope to see them.

\subsection{Amplitude and phase stability}

If pulsating sdB stars are observed over an extended time period,
it is common to detect amplitude variability in many, if not all, of the
pulsation frequencies \citep[e.g. ][]{simon02,reed07b,zhou}. 
Variations on time scales of days to weeks sometimes
appear in clearly resolved pulsation spectra where
mode beating cannot be the cause \citep{reed06a}.
Given the limitations of ground--based observations, in most cases long-term variability cannot 
be securely attributed to beating between pulsations, and could also result from intrinsic amplitude or phase modulations.

\begin{figure}
\includegraphics[width=8.4cm]{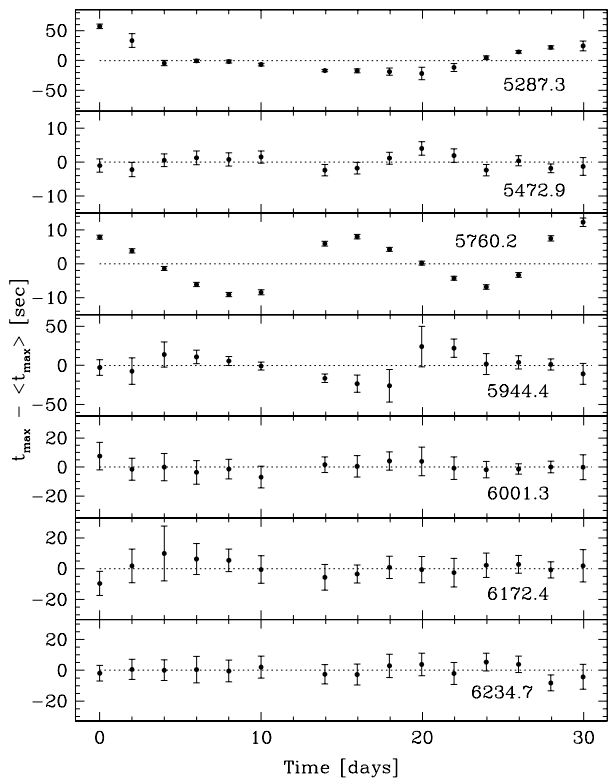}

\caption{Pulsation phases for 7 well-separated strong pulsations seen in  KIC\,010139564.  Here, we used a least-squares fit to the dominant frequency in each band (with fixed frequency) and fit 2-day subsets of the data to obtain the pulsation phases and amplitudes (see Fig. \ref{fig05}).  The phase is expressed in terms of the time of the first maximum (in seconds) after the start of the observations.  The three highest-frequency periodicities (bottom panels) show no phase variation over the course of the run, but the four lower frequency periodicities show apparent variations.}\label{fig04}
\end{figure}

\begin{figure}

\includegraphics[width=84mm]{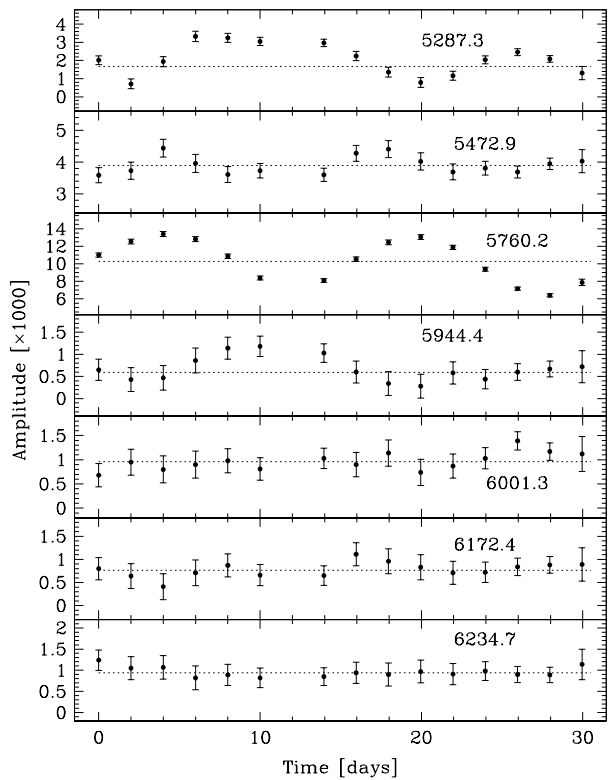}
\caption{Pulsation amplitudes (in mma) for the same 7 well-separated strong pulsations used in Fig.~\ref{fig04}, computed using the same procedure.  The three highest-frequency periodicities (bottom panels) show no obvious amplitude variation over the course of the run, but the four lower frequency periodicities show apparent variations.}\label{fig05}
\end{figure}

As previously mentioned, a true strength of \emph{Kepler}
data is that we have nearly continuous, single-instrument data
which can be examined for amplitude and phase stability. We broke the data up into six 5-day spans, and computed an averaged temporal spectrum of these six shorter runs to obtain estimates of the position of the centroid of pulsation ``power'' at each of 7 cleanly separated regions in the averaged temporal spectrum.  We then used those 7 frequencies (flagged with asterisks in Table~\ref{tab01}) as fixed quantities to examine them for phase and amplitude stability over two--day spans of data.  We used a fixed value for each of these 7 frequencies (see Table~\ref{tab01}) and then did a non-linear least squares (NLLS) fit to the data in 2-day intervals to find the best-fitting values for their amplitudes and phases.
The results are shown in Fig.~\ref{fig04} and Fig.~\ref{fig05}.


Breaking the data into two-day chunks reduces the frequency
resolution, but it allows us to track the amplitude and phase of the
variation with time.  If the amplitude is modulated because of
beating between two coherent modes, then the phase of the light curve
should flip sign at the time that the amplitude reaches a minimum (or
maximum).  Both phase and amplitude curves should look sinusoidal
as well, as if there are only two modes beating.
If the amplitude or phase curves show non-sinusoidal variations on time scales shorter than the run length, and the extrema in the amplitude variations do not occur at zero--crossings of the phase variations, then
simple beating between two or three modes is not taking place.

From Fig.~\ref{fig04} we see that the three higher frequencies
examined clearly demonstrate phase stability over these time-scales. Their amplitudes are also stable (Fig.~\ref{fig05}).
Over the course of \emph{Kepler's}
three and a half year mission, these frequencies will be useful for
searching for secular period changes caused by the evolution of the stars, and may be useful for precise timing to search for sub--stellar companions \citep{silv07}.

The four lower frequency periodicities (top 4 panels) show 
phase and amplitude variations over the 30.5 days of data.  Of those, two (5472.9~$\mu$Hz, f7 and 5760.2~$\mu$Hz, f8) 
vary on time scales shorter than
the length of the observing run.  Their phase and amplitude variations are out of phase by about 90 degrees, as one would expect from beating between closely spaced stable oscillation 
frequencies.  Thus we conclude that these two frequencies appear to be (nearly) resolved into two or three stable periodicities.  The largest amplitude 
periodicity (5760.2~$\mu$Hz, f8) 
shows sinusoidal amplitude and phase variations, and is
cleanly resolved in the temporal spectrum analysis (see Table~\ref{tab01} and Fig.~\ref{fig02}) with separations of approximately 0.75~$\mu$Hz.  The 5472.9~$\mu$Hz region can also be fit by 
three periodicities spanning 1.7~$\mu$Hz, but not with the same level of confidence -- note also that two of those may be linear combination frequencies.

The peak at 5944.4~$\mu$Hz (f10) also shows variation in amplitude in Fig.~\ref{fig05} but the phase variation shows a less regular pattern that does not match the expected phasing.  In part this may be a result of the pulsation amplitude reaching small values (i.e. around day 20).  The power spectrum analysis suggests that this is a pair of peaks separated by approximately 0.61~$\mu$Hz, but it is not well resolved, and again may include a periodicity at a linear combination frequency.

The spacings hinted at above, based not on the least-squares fits in Table~\ref{tab01} but on the beating pattern, are all around 0.6 to 0.8~$\mu$Hz - roughly twice the frequency resolution of the data. 
If we attribute these splittings to rotation, then we can suggest a tentative value for the rotation period of approximately 16-20 days for  KIC\,010139564.  
Clearly a longer run is needed to firmly resolve these peaks and confirm that they are indeed multiplets.

Given the current data, the
best case we find for a periodicity that has an intrinsic variation is the region around 5287.3~$\mu$Hz.  
This shows phase variation 
on a time scale longer than the span of observation. The amplitude varies quasiperiodically with a period roughly 0.75 times the observation span, but on a different (shorter) time scale than the phase variation.  In the temporal spectrum, this is a region 
where NLLS fitting failed, and residual power remained after prewhitening.  Given the phase and amplitude variations seen in the data, we suspect that this mode shows intrinsic amplitude and phase instability, unlike the other strong periodicities in this star.

\subsection{The low--frequency oscillation}
The periodicity labelled f0 in Table \ref{tab01} is notable for being in the frequency range observed in the V1093\,Her
stars, and near the low-frequency periodicities in the hybrid (DW~Lyn) pulsators.  This single periodicity, with an amplitude of 0.45~mma is the smallest-amplitude variation seen in this frequency range for hybrid subdwarf B pulsators.  In the other members of this class, the amplitude ratio between the short and long period modes is in the range from 5 to 20 or more.  In this star, the ratio is a factor of 23, which is consistent with that seen in the known hybrid pulsators.

\begin{figure*}
\begin{minipage}{126mm}
\includegraphics[width=126mm]{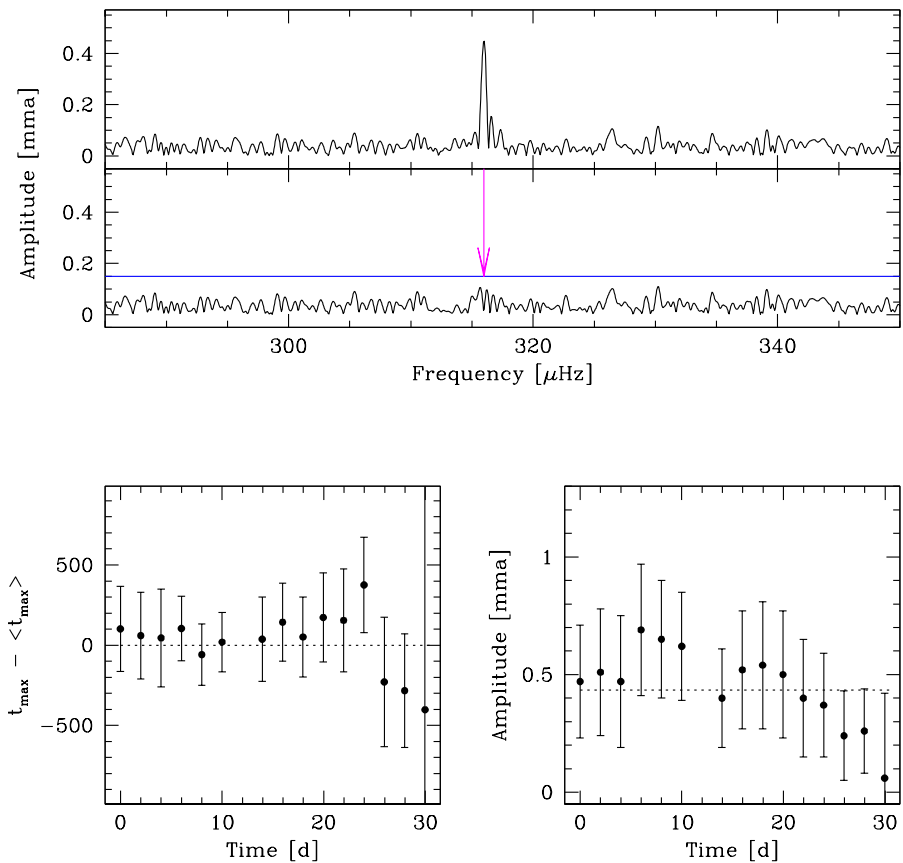}
\caption{Top panel: Temporal spectrum at low frequencies showing the f0 periodicity.  Lower panels: Time variation of the amplitude and phase of the f0 periodicity, 
computed as in Fig.~\ref{fig04} and Fig.~\ref{fig05}.
}\label{fig06}
\end{minipage}
\end{figure*}

Fig. \ref{fig06} shows the temporal spectrum in the low frequency range surrounding f0, along with the time dependence of the pulsation amplitude and phase 
(computed in the same manner as for the short period modes).
A few things are notable about this mode.  First, it seems to be a single oscillation mode - fitting a single sinusoid removes the temporal spectrum peak and leaves no residuals that can be distinguished from noise.  It remains visible through the entire 30.5 day interval, and maintains phase (within the errors).  However, there is a
possible
trend towards decreasing amplitude over most of the observing interval.  With further monitoring, we hope to determine if this mode will remain visible.

Interestingly, with the spectroscopically determined $T_{\rm eff}$ of 32,500K, KIC\,010139564 is significantly hotter than the other hybrid pulsators.
Finally, we add that there is a slim chance that this periodicity may 
be contributed by the contaminating source -- if Nature is being perverse.

\section{Conclusions and Future Work}

We present results from a study of the first \emph{Kepler} data on a newly-discovered pulsating subdwarf B star that shows mostly short-period pulsations characteristic of the V361~Hya stars, along with a single oscillation at a frequency characterstic of the V1093\,Her stars.  Thus KIC\,010139564 may be a member of the hybrid sdB pulsator class. The \emph{Kepler} data have realized the potential to obtain a nearly uninterrupted time series of photometric data for this type of pulsator that
spans over 15,000 pulsation cycles. Previous ground-based efforts have demonstrated that such
data are essential for overcoming aliasing issues, caused
by gaps and the necessity of using multiple instruments on varied
apertured telescopes from the ground.  The science questions raised by ground-based attempts at photometry of these stars can be addressed directly with {\em Kepler} data.

These data have allowed us to confidently detect 21 periodicities, a few of which may be linear combination frequencies.  At least one of the main periodicities remains unresolved, 
but it is not yet clear whether we are seeing beating or
stochasitic variations.
Residual power indicates the presence of possible pulsations below the current
detection limit. As such, with longer duration \emph{Kepler} observations,
we can anticipate finding more periodicities at even lower amplitudes.  
{\em Kepler} can place much more stringent upper limits on pulsation amplitudes in these stars than we can achieve from ground-based data, so we can also hope to determine if ground-based
observations of these stars that reveal simple pulsators have been limited by a higher detection
threshold. 
We also anticipate resolving rotationally-split
multiplets, which will add stringent observational constraints to
theoretical models.

The data presented here confirms the potential of \emph{Kepler} for
sdB asteroseismology. KIC\,010139564 has a frequency spectrum
that is as rich as (or richer than) possible for low-degree modes only, and there is still more to come. 
We anticipate further, longer-duration observation during the second
year of operations, when \emph{Kepler} will no longer be in survey
mode.

\section*{Acknowledgments}
For R.H.{\O}. and S.B., the research leading to these results has received funding from the European
Research Council under the European Community's Seventh Framework Programme
(FP7/2007--2013)/ERC grant agreement n$^\circ$227224 (PROSPERITY) and
from the
Research Council of K.U.Leuven (GOA/2008/04). ACQ is supported by the Missouri Space Grant Consortium, funded by NASA.  A.B. acknowledges support from the Polish Ministry of Science (554/MOB/2009/0). 
Funding for this Discovery mission is provided by NASA's Science Mission
Directorate. The authors gratefully acknowledge the entire \emph{Kepler}
team, whose efforts have made these results possible.

\label{lastpage}

\end{document}